\documentclass{iau} 
\usepackage{graphicx}
\usepackage{amssymb, amsmath}
\usepackage{color}
\usepackage[colorlinks=true, linkcolor=blue, citecolor=blue, urlcolor=blue]{hyperref}
\usepackage{natbib}
\newcommand{\aap}{\textit{A\&A}}
\newcommand{\apj}{\textit{ApJ}}
\newcommand{\apjs}{\textit{ApJS}}
\newcommand{\araa}{\textit{ARA\&A}}
\newcommand{\mnras}{\textit{MNRAS}}

\title[Panchromatic SED fitting codes and modelling techniques] 
{Panchromatic SED fitting codes \\and modelling techniques}

\author[Maarten Baes]   
{Maarten Baes}

\affiliation{Sterrenkundig Observatorium, Universiteit Gent, \\ Krijgslaan 281 S9, 9000 Gent, Belgium 
\\ email: {\tt maarten.baes@ugent.be}}

\pubyear{2019}
\volume{341}  
\jname{PanModel2018: Challenges in Panchromatic Galaxy Modelling with Next Generation Facilities}
\editors{M. Boquien, E. Lusso, C. Gruppioni, \& P. Tissera, eds.}
\begin{document}

\maketitle

\begin{abstract}
Modelling and interpreting the SEDs of galaxies has become one of the key tools at the disposal of extragalactic astronomers. Ideally, we could hope that, through a detailed study of its SED, we can infer the correct physical properties and the evolutionary history of a galaxy. In the past decade, panchromatic SED fitting, i.e.\ modelling the SED over the entire UV--submm wavelength regime, has seen an enormous advance. Several advanced new codes have been developed, nearly all based on Bayesian inference modelling. In this review, we briefly touch upon the different ingredients necessary for panchromatic SED modelling, and discuss the methodology and some important aspects of Bayesian SED modelling. The current uncertainties and limitations of panchromatic SED modelling are discussed, and we explore some avenues how the models and techniques can potentially be improved in the near future. 
\keywords{galaxies: stellar content; galaxies: ISM; infrared: galaxies; ultraviolet: galaxies; methods: statistical; dust, extinction; radiative transfer}
\end{abstract}

\firstsection
\section{Introduction}
\vspace*{0.5em}

\noindent The main goal of extragalactic astronomy is understanding the formation and evolution of galaxies throughout cosmic history. The global spectral energy distributions (SEDs) of galaxies are a key element in this grand challenge. Indeed, the different astrophysical processes that shape a galaxy, such as star formation, stellar evolution, chemical enrichment and processes in the interstellar medium, all leave an imprint on the SED. Ideally, we could hope that, through a detailed study of its SED, we can infer the physical properties and the evolutionary history of a galaxy. This detailed study is commonly referred to as SED modelling or SED fitting. For a general background on galaxy SED modelling we refer to the excellent reviews on this topic by \citet{1980FCPh....5..287T}, \citet{2011Ap&SS.331....1W} and \citet{2013ARA&A..51..393C}.

Conceptually, SED modelling is relatively straightforward, and usually is done in two phases. The first phase consists of the construction of a theoretical or empirical parameterised SED model that is assumed to incorporate all the physical ingredients and processes that shape the SEDs of galaxies. The basis of all galaxy SED models are simple stellar populations (SSPs), which  describe the SED of a coeval stellar population at a given age, metallicity and abundance pattern. Under the hood, any SSP model requires assumptions on the initial mass function, stellar evolution, and stellar spectral libraries, and different sets of SSP models are available in the literature \citep[e.g.,][]{1993ApJ...405..538B, 2003MNRAS.344.1000B, 1999ApJS..123....3L, 2005MNRAS.362..799M, 2009ApJ...699..486C}. From a given SSP library, a composite stellar population is built by combining different SSPs according to a chemical enrichment history (CEH) and star formation history (SFH). In practice, a parameterised galaxy SED model usually consists of a huge grid of individual SED templates, each of them corresponding to a specific set of values for the model parameters.

The second phase in the SED modelling consists of determining the parameter values of the model SED that reproduces the observed galaxy SED best. Again, this sounds relatively straightforward, but it is more complex in practice. Firstly, the theoretical model SEDs need to be converted to mock SED data that can immediately be compared to the observations, which involves redshifting, filtering through the spectral response curves of the instruments, and spatial redistribution over the point spread function of the instruments \citep{2011Ap&SS.331....1W}. Secondly, determining which model SED reproduces the observed SED ``best'' is not a straightforward task with a unique answer. As for most optimisation problems, the simplest answer is usually some form of $\chi^2$ optimisation, but this is not necessarily the ``best'' answer, especially when the data are noisy.

Until fairly recently, the observed SEDs of galaxies were often limited to the UVOIR (UV, optical and NIR) regime. At these wavelengths, the SED is dominated by stellar continuum emission, with usually minor contributions from ionised gas or an active galactic nucleus. The UVOIR SED could hence be modelled as a weighted sum of SSP SEDs, except for the disturbances caused by interstellar dust. Dust grains in the interstellar medium of galaxies are extremely efficient at absorbing and scattering UV and optical radiation. The net effect of dust absorption and scattering on the SED of galaxies is labelled attenuation. Correcting UVOIR galaxy SEDs for the effects of attenuation is nontrivial, as the shape of the attenuation curve depends on both the optical properties of the dust and the relative star-dust geometry \citep{2001MNRAS.326..733B, 2004A&A...419..821T, 2017MNRAS.470..771T, 2018ApJ...869...70N}. Nearly all SED modelling codes assume a parameterised attenuation curve \citep[e.g.,][]{2000ApJ...533..682C, 2000ApJ...539..718C, 2017MNRAS.472.1372L, 2018A&A...619A.135B}.

The effect of dust attenuation on the SED of galaxies, and thus on the physical properties derived from SED fitting, are substantial. In nearby, star-forming galaxies, interstellar dust grains absorb about 30\% of the radiative energy emitted by stars \citep{2002MNRAS.335L..41P, 2016A&A...586A..13V, 2018A&A...620A.112B}. This absorbed energy is re-emitted at longer wavelengths, i.e., in the mid-infrared, far-infrared, and submm regimes. Except for some extreme cases where synchrotron emission from an AGN is important \citep[e.g.,][]{2010A&A...518L..53B}, thermal dust emission is the dominant contributor to the SED of galaxies in the infrared regime. In the past two decades, space missions such as WISE, Spitzer, and Herschel have opened up these regimes, and have observed large areas of the extragalactic sky. Major efforts have been undertaken to create databases with observed panchromatic SEDs\footnote{In this review we concentrate on the wavelength regime from UV to submm wavelengths, and we leave out the gamma-ray, X-ray, microwave and radio regimes. The UV--submm regime is reflects nearly all of the energy emitted by stars, either directly or reprocessed by dust absorption and re-emission. Outside this regime, non-stellar processes dominate the emission.} for thousands of nearby and distant galaxies \citep[e.g.,][]{2007ApJS..172....1S, 2012MNRAS.427..703S, 2015MNRAS.452.2087L, 2016ASSP...42...71V, 2017MNRAS.464.1569A, 2018A&A...609A..37C}. This availability of data across the entire UV--submm spectrum gives us much stronger constraints on both the stellar populations and the interstellar dust content in galaxies, and it is hence wise to use all of this information in the interpretation of galaxy SEDs. On the other hand, self-consistently modelling the entire UV--submm SED of galaxies, i.e., panchromatic SED modelling, also brings along some intellectual and technical challenges. 

\section{Panchromatic SED fitting codes}
\vspace*{0.5em}

\noindent At UVOIR wavelengths, the SED of a galaxy can be regarded as the weighted sum of SEDs corresponding to different SSPs, with an additional attenuation correction factor. SED fitting then comes down to determining the optimal weights, which is a linear inversion problem that can be solved by classical inversion techniques. This approach is used by many spectral fitting codes including STARLIGHT \citep{2005MNRAS.358..363C}, STECKMAP \citep{2006MNRAS.365...74O}, VESPA \citep{2007MNRAS.381.1252T}, ULySS \citep{2009A&A...501.1269K}, and FIREFLY \citep{2017MNRAS.472.4297W}. When we consider the full panchromatic SED of a galaxy, inversion techniques are not suitable anymore, because the SED is no longer a simple weighted sum of components when thermal dust emission is taken into account. Moreover, the number of free parameters to be considered is typically rather large, with many possible degeneracies between the parameters. 

Nearly all panchromatic SED fitting codes operational today are based on Bayesian inference modelling. Bayes' theorem can be written as
\begin{equation}
P({\mathbf{\Theta}}|{\mathbf{D}},{\text{M}}) 
= 
\frac{
P({\mathbf{\Theta}}|{\text{M}})\,
P({\mathbf{D}}|{\mathbf{\Theta}},{\text{M}})
}{
P({\mathbf{D}}|{\text{M}})
}
\end{equation}
In this equation, ${\mathbf{\Theta}}$ corresponds to a set of parameters of the theoretical SED model M, and ${\mathbf{D}}$ is the observed SED data set. In words, Bayes' theorem states that the posterior distribution of the parameters ${\mathbf{\Theta}}$ can be found as the normalised product of the prior distribution of ${\mathbf{\Theta}}$ and the likelihood of the data, given this set of parameters. This formula enables us to draw rigorous constraints on the model parameters based on the observed SED, within the modelling framework M, without any specific assumptions about the shape of the posterior probability distribution. To determine the ``best'' parameters for a specific data set, the posterior distribution is marginalised over all parameters except one, which results in a series of one-dimensional posterior probability density functions,
\begin{equation}
P(\Theta_i | {\mathbf{D}},{\text{M}}) 
= 
\int P(\Theta_i | {\mathbf{D}},{\text{M}})\, 
{\text{d}}\Theta_1\cdots{\text{d}}\Theta_{i-1}\,{\text{d}}\Theta_{i+1}\cdots{\text{d}}\Theta_m
\end{equation}
For a given parameter, the median/mean and standard deviation of this marginalised distribution are taken as the best estimate for this parameter and the associated uncertainty. Specifically interesting about Bayesian inference modelling in the context of SED fitting is that instrumental systematics can be incorporated in the modelling in a consistent way. Also, full posterior distributions for the observables are available, and complex correlations between model parameters can be investigated. For a more detailed description of Bayesian inference in the context of SED modelling, we refer to \citet{2003MNRAS.341...33K}, \citet{2014ApJS..215....2H} and \citet{2016MNRAS.462.1415C}.

Two aspects in Bayesian SED fitting deserve some special attention. The first one is the need for calculating the marginalised probability density functions for all parameters, which involves complex multi-dimensional integrations over the parameter space. The easiest approach to deal with this problem is to use a regular model grid, which makes the marginalisation rather trivial. Obvious disadvantages of a grid approach are that this method can become inefficient and very memory-intensive, in particular when many different parameters need to be considered. In this case, the grid often needs to be made rather coarse, which can result in poor accuracy and can lead to finite-resolution effects \citep[see e.g.,][]{2019arXiv190305933N, 2019arXiv190306715D}. Alternative sampling methods such as Markov Chain Monte Carlo or nested sampling can avoid these problems. These methods result in an increase in the efficiency and reliability of the computation of the marginalised distributions, as calculations in irrelevant parts of the parameter space are avoided, and the computational effort is concentrated on the important parts only \citep{2009ApJ...699..486C, 2011ApJ...737...47A}. A challenge for such methods is that the theoretical SED needs to computable for {\em{any}} set of values of the parameters, which usually requires some kind of interpolation. Artificial neural networks are a promising technique to do this interpolation \citep{2014ApJS..215....2H}.

A second important aspect is the choice of the model M and the prior distribution $P({\mathbf{\Theta}}|{\text{M}})$, which reflects our assumed knowledge about the model parameters without any knowledge of the particular data set ${\mathbf{D}}$. In some cases, the likelihood distribution will be informative enough, such that the prior distribution has only a negligible impact on the resulting posterior distribution. In other cases, however, the data do not really constrain some model parameters, and the choice of the priors (and more generally, the choice of the model) can have a significant effect on the final results. Standard choices for the priors are often flat distributions in linear or logarithmic space, although it is clear that suitable priors should be preferred if possible. One advanced approach to obtain realistic priors is so-called hierarchical Bayesian SED modelling, which can be applied when modelling a sample of galaxies. In this case, the prior distributions are inferred from the data. This is achieved by parameterising the shape and position of the new prior distribution with hyperparameters, which control the distribution of the parameters \citep{2012ApJ...752...55K, 2018MNRAS.476.1445G}. 

Over the past few years, many different Bayesian SED fitting codes have been conceived, and some of them were specifically designed to cover the entire UV--submm wavelength range. Two pioneering panchromatic SED fitting codes are MAGPHYS \citep{2008MNRAS.388.1595D, 2015ApJ...806..110D} and CIGALE \citep{2009A&A...507.1793N, 2019A&A...622A.103B}. These codes are grid-based Bayesian codes that fit the entire UV--submm SED of galaxies under the constraint that the energy absorbed by dust is balanced by the energy emitted at long wavelengths. Both codes have been applied extensively in the literature \citep[e.g.][]{2012MNRAS.427..703S, 2018MNRAS.475.2891D, 2018A&A...620A..50M}. More recent codes that include advanced options on the prior distributions and/or the sampling methodology include BayeSED \citep{2014ApJS..215....2H}, BEAGLE \citep{2016MNRAS.462.1415C}, Prospector \citep{2017ApJ...837..170L} and BAGPIPES \citep{2018MNRAS.480.4379C}. Comprehensive comparisons between different panchromatic galaxy SED codes have been presented by \citet{2016A&A...589A..11P}, \citet{2017ApJ...837..170L} and \citet{2019A&A...621A..51H}.

\section{Uncertainties in panchromatic SED models}
\vspace*{0.5em}

\noindent In spite of the progress that has been made over the past few years, there are still substantial uncertainties linked to SED modelling. In fact, all the different inputs in the SED modelling processes are prone to at least some degree of uncertainty. 

In a series of papers, \citet{2009ApJ...699..486C, 2010ApJ...708...58C} and \citet{2010ApJ...712..833C} discuss how uncertainties in stellar evolution and the IMF propagate to uncertainties in SSP models \citep[see also][]{2011Ap&SS.331....1W}. Among the most important uncertainties are particular late phases in stellar evolution (thermally pulsating AGB stars, blue stragglers, etc.), the effects of binary stellar evolution, and the high-mass slope of the IMF. They warn that the interpretation of the resulting uncertainties in the derived colours is highly non-trivial because many of the uncertainties are likely systematic, and possibly correlated with the physical properties of galaxies. As an indication, they find that stellar masses in the local Universe typically carry errors of  about 0.3 dex with little dependence on luminosity or color, while this uncertainty increases to about 0.6 dex for bright red galaxies at $z\sim2$.

Other components in panchromatic SED modelling also carry their uncertainties, and propagate them to the final model SEDs. The treatment of dust in particular is prone to many systematic uncertainties. The shape of the attenuation curve is one of these; most codes use either the empirical \citet{2000ApJ...533..682C} or \citet{2000ApJ...539..718C} prescriptions for the attenuation curve. Whether these prescriptions are realistic is subject to quite some debate, and it is definitely difficult to constrain the attenuation based on UVOIR data alone. In principle, strong constraints on the dusty ISM can be obtained from the infrared portion of the SED, and this information can be coupled to the attenuation obtained at UVOIR wavelengths. The relation between the amount of dust, the grain properties, and the attenuation is far from trivial, however, and strongly depends on the star-dust geometry, the structure of the ISM, and the importance of scattering. Further complications arise from the fact that galaxy SEDs are anisotropic at short wavelengths (more attenuation for higher inclinations), whereas they are largely isotropic in the FIR/submm regime. This breaks the energy balance approach. Moreover, the actual physical dust models are subject to substantial uncertainties \citep[][and references therein]{2017A&A...602A..46J, 2018ARA&A..56..673G}, and the implicit assumption that the dust properties are uniform across a galaxy is most probably a major simplification \citep{2012ApJ...756...40S, 2012MNRAS.423...38M, 2018A&A...613A..43R}.

Finally, significant uncertainties in SED models are linked to how the different SSP and dust models are combined to a global SED model, and how this model is fitted to the data. As already indicated, most SED modelling codes are Bayesian in nature, and the choice of the model and the priors is crucial. A typical example is the choice of the SFH in the model. In many codes, the SFH is parameterised using simple analytical prescriptions \citep[e.g.,][]{2009ApJS..184..100L, 2012MNRAS.422.3285P, 2016A&A...585A..43C}, which rule out a significant fraction of the full diversity of SFH shapes a priori. Using the Prospector code equipped with a non-parametric SFH module, \citet{2018arXiv181103637L} demonstrate that the choice of prior rather than the photometric noise is the primary determinant of the size and shape of the posterior SFH. A similar, but probably less dramatic, effect is to be expected for the CEH. While Bayesian SED fitting codes can, in principle, include any parametric or non-parametric CEH, most codes assume a single metallicity for the entire composite population. The impact of this simplification has not been extensively explored \citep{2013ARA&A..51..393C}.

Given all these uncertainties, simplifications and potential systematics, one could wonder whether we should believe the results of panchromatic SED modelling efforts. One way to answer this question, at least partially, is by applying panchromatic SED modelling to mock observables from hydrodynamically simulated galaxies. In a series of papers on this topic, \citet{2015MNRAS.446.1512H} and \citet{2015MNRAS.453.1597S, 2018MNRAS.476.1705S} apply MAGPHYS to a suite of high-resolution hydrodynamical simulations. They conclude that most physical parameters are recovered fairly well in most cases, albeit with a number of caveats and systematics. In particular, recovering the SFH proved to be difficult, even though the mock photometry was fit reasonably well. 

\section{Towards more advanced SED templates}
\vspace*{0.5em}

\noindent In order to bring panchromatic SED modelling to a new level, it is clear that more work is needed to decrease the uncertainties on all of these different aspects: stellar evolution, SSP models, dust models, attenuation curves, SFHs, CEHs, realistic priors, etc. 

An obvious way to improve the handling of dust attenuation and emission is dust radiative transfer modelling. Only when dust radiative transfer is rigorously calculated in a realistic framework, one can be sure that attenuation and emission by dust are consistent, i.e., the dust energy balance is satisfied on global and local scales. The field of 3D dust radiative transfer has advanced enormously in recent years \citep[for an overview, see][]{2013ARA&A..51...63S}. Most progress has been made on Monte Carlo radiative transfer techniques, with novel acceleration techniques \citep{2011ApJS..196...22B, 2016A&A...590A..55B}, advanced grids and associated traversal algorithms \citep{2001A&A...379..336K, 2013A&A...560A..35C, 2013A&A...554A..10S, 2014A&A...561A..77S}, and hybrid parallelisation strategies \citep{2017A&C....20...16V}. As a result, grids of self-consistent radiative transfer models for AGNs \citep{2012MNRAS.420.2756S, 2016MNRAS.458.2288S, 2015A&A...583A.120S} and galaxies \citep{2000MNRAS.313..734E, 2018ApJS..236...32L} have appeared in the literature, which can be used as the base components in SED modelling efforts.

These efforts could be taken a step further through the combination of high-resolution hydrodynamical modelling and 3D dust radiative transfer. Cosmological hydrodynamical simulations have only fairly recently achieved sufficient realism and statistics to provide a predictive model for the Universe \citep{2014MNRAS.444.1518V, 2015MNRAS.446..521S, 2018MNRAS.473.4077P}, and zoom simulations have managed to produce high-resolution galaxy models that reproduce the properties and scaling relations of galaxies in many aspects \citep{2015MNRAS.454...83W, 2017MNRAS.467..179G}. Combined with the advances in 3D dust radiative transfer modelling, it is now possible to self-consistently calculate panchromatic mock observables for hydrodynamically simulated galaxies \citep{2010MNRAS.403...17J, 2011ApJ...743..159H, 2014MNRAS.439.3868D, 2015A&A...576A..31S, 2017MNRAS.469.3775G, 2019MNRAS.483.4140R}. 

Interestingly, 3D dust radiative transfer postprocessing can now be applied to entire populations of galaxies from cosmological simulations. \citet{2016MNRAS.462.1057C, 2018ApJS..234...20C} applied the SKIRT radiative transfer code \citep{2015A&C.....9...20C} to almost half a million simulated galaxies from the suite of EAGLE simulations \citep{2015MNRAS.446..521S}. Such collections of panchromatic mock SEDs could form an ideal set of templates for panchromatic SED modelling. Indeed, for these templates, attenuation and thermal emission by dust is calculated in a self-consistent way, and in a realistic 3D geometry. Futhermore, the SFHs and CEHs of simulated galaxies are probably more realistic than the simple analytical prescriptions used in most panchromatic SED fitting codes. Finally, the current generation of cosmological hydrodynamical simulations manage to reproduce many observed population characteristics (stellar mass function, SFR function, etc.) and scaling laws (mass-metallicity relation, relation between stellar mass and angular momentum, etc.), such that the population of galaxies in a cosmological simulation could in principle be used to construct more informative priors on the model parameters. 

\section{Conclusions}
\vspace*{0.5em}

\noindent Panchromatic UV--submm SED fitting has received an enormous boost in the past decade. In their 2011 review, \citet{2011Ap&SS.331....1W} concluded that most panchromatic SED fitting codes are still in their testing phase, and argued that the large number of derived parameters and our still limited knowledge of their respective degeneracies and systematic uncertainties make it difficult to go a step further and fully use the full power of SED fitting. Since these conclusions were written, several new codes, all based on Bayesian inference modelling, have seen the light or have been profoundly upgraded, and our knowledge on the importance of realistic priors and noise models, and our methods to explore the parameter space and calculate posterior distributions have improved substantially. Without a doubt, there are still many uncertainties on many aspects involved in the SED modelling process, but these are gradually better understood, and their impact on the derived properties of galaxies are being investigated in detail. The current generation of cosmological hydrodynamical models, combined with 3D dust radiative transfer techniques, could be an interesting path to bring panchromatic modelling to a new level. In short: the future of panchromatic SED modelling is bright. 

\section*{Acknowledgements}
\vspace*{0.5em}

\noindent The author thanks the organisers of the PanModel2018 conference for a very interesting and entertaining conference, and particularly for the invitation to deliver an invited review on panchromatic SED fitting codes and modelling techniques. It was a honour to be asked. Wouter Dobbels and Peter Camps are acknowledged for comments on this manuscript.

\begin{discussion}
\discuss{Fabio Fontanot}{How much can we hope to reduce the inferred error on recovered physical parameters obtained from panchromatic SED fitting?}

\discuss{Maarten Baes}{This is very hard to estimate at this moment. We know from various studies that the choice of the prior is fundamental, and currently it seems unrealistic to claim uncertainties smaller than $\sim$0.3 dex on stellar masses. Star formation histories are much more uncertain, and it would be interesting to test whether it is possible to reduce these uncertainties by building in priors as obtained from cosmological hydrodynamical simulations.}

\discuss{Alexa Villaume}{Cosmological hydrodynamical simulations rely on subgrid prescriptions to include star formation. Is there something circular to using those to make templates for SED fitting to obtain information about star formation?}

\discuss{Maarten Baes}{I believe that, indeed, there is some kind of circularity in this process, but this is unavoidable as long as we cannot model all processes in galaxy evolution ab initio. For most cosmological hydrodynamical simulations, the parameters of the subgrid models are calibrated based on observed population properties and scaling relations in the local Universe, which does not really put very strong constraints on the possible range of star formation histories. In general, the range of SFHs obtained from cosmological hydrodynamical simulations is broader than the set of parameterised SFHs often used in SED modelling.} 

\discuss{Lapo Fanciullo}{How is dust formation and evolution modelled in cosmological hydrodynamical simulations? What are the properties of the dust in the simulation?}

\discuss{Maarten Baes}{In our SKIRT-EAGLE work \citep{2016MNRAS.462.1057C, 2018ApJS..234...20C, 2017MNRAS.470..771T, 2019MNRAS.484.4069B} we have adopted a fixed grain model in our post-processing radiative transfer modelling. More particularly, we have used the BARE\_GR\_S model from \citet{2004ApJS..152..211Z}, and assumed a fixed dust-to-metal fraction of 30\% for the diffuse cold gas. In principle, it is possible to include a dust evolution model into a cosmological hydrodynamical model \citep[e.g.,][]{2015MNRAS.449.1625B, 2017MNRAS.468.1505M, 2018MNRAS.478.2851M, 2018MNRAS.478.4905A}. Unfortunately, the detailed physics involved in the different processes involved, like grain formation, coagulation, shattering, sputtering and drag forces, are still poorly constrained, and the coupling to a dust evolution model makes the hydrodynamical simulations extremely computationally demanding.}

\discuss{Hiroyuki Hirashita}{\citet{2018MNRAS.478.4905A} also have a cosmological simulation with dust evolution, and with a grain size evolution. How good should the spatial resolution be to obtain a reliable SED?}

\discuss{Maarten Baes}{In \citet{2018ApJS..234...20C} we have experimented with different models, and we found that a minimum of about 250 dust particles per galaxy in the simulation is needed for the dust radiative transfer to yield meaningful results. Obviously, the higher the spatial/mass resolution, the better.}

\end{discussion}

\end{document}